\documentclass[12pt,fleqn]{article}
\usepackage{graphicx}

\def\be{\begin{equation}}
\def\ee{\end{equation}}
\def\bea{\begin{eqnarray}}
\def\eea{\end{eqnarray}}
\def\nnb{\nonumber}
\def\bbuildrel#1_#2^#3{\mathrel{\mathop{\kern 0pt#1}\limits_{#2}^{#3}}}
\def\slash#1{\setbox0=\hbox{$#1$}#1\hskip-\wd0\dimen0=5pt\advance
       \dimen0 by-\ht0\advance\dimen0 by\dp0\lower0.5\dimen0\hbox
         to\wd0{\hss\sl/\/\hss}}

\newcommand{\TT}{\rule[-2mm]{0mm}{7mm}}
\newcommand{\scs}{\scriptscriptstyle}
\newcommand{\f}{\frac}
\newcommand{\fm}[2]{{\textstyle \frac{#1}{#2}}}
\newcommand{\al}{\alpha_s}
\newcommand{\wi}{\widetilde}
\newcommand{\e}{\epsilon}
\newcommand{\ve}{\varepsilon}

\newcommand{\me}[1]{\langle#1\rangle}
\newcommand{\cd}{\!\cdot\!}

\newcommand{\newsection}[1]{\section{#1}\setcounter{equation}{0}}

\begin{document}
\begin{titlepage}

\begin{flushright}
  {\bf TUM-HEP-455/02\\
       Alberta Thy 05-02\\
       IFT-5/2002\\
       hep-ph/0203135}\\[2cm]  
\end{flushright}

\begin{center}

\setlength {\baselineskip}{0.3in} 
{\bf\Large Completing the NLO QCD calculation of $\bar{B} \to X_s \gamma$}\\[2cm]

\setlength {\baselineskip}{0.2in}
{\large  Andrzej J. Buras$^{^{1}}$, Andrzej Czarnecki$^{^{2}}$, 
Miko{\l}aj Misiak$^{^{3}}$ and J\"org Urban$^{^{1}}$}\\[5mm]

$^{^{1}}${\it Physik Department, Technische Universit\"at M\"unchen,\\
               D-85748 Garching, Germany}\\[3mm]

$^{^{2}}${\it Department of Physics, University of Alberta,\\
               Edmonton, Alberta, Canada T6G 2J1}\\[3mm]

$^{^{3}}${\it Institute of Theoretical Physics, Warsaw University,\\
                 Ho\.za 69, PL-00-681 Warsaw, Poland}\\[2cm] 

{\bf Abstract}\\
\end{center} 
\setlength{\baselineskip}{0.2in} 

We evaluate two-loop $b\to s\gamma$ matrix elements of all the
four-quark operators containing no derivatives. Contrary to previous
calculations, no expansion in the mass ratio $m_c/m_b$ is performed,
and all the possible Dirac and flavor structures are included.
Consequently, we are able to provide the last item in the NLO analysis
of $\bar{B} \to X_s \gamma$ that has been missing so far, namely the
two-loop matrix elements of the QCD-penguin operators.  Due to 
smallness of the Wilson coefficients of those operators in the
Standard Model, their effect on the branching ratio is small: a
reduction by roughly $1\%$. We find
BR$[\bar{B} \to X_s \gamma]_{E_{\gamma} > 1.6~{\rm GeV}} 
= (3.57 \pm 0.30)\times 10^{-4}$. 

\end{titlepage} 
\setlength{\baselineskip}{0.23in}

\newsection{Introduction}
\label{sec:intro}

The decay $\bar{B} \to X_s \gamma$ constitutes a stringent test of the
Standard Model (SM) and many of its extensions. In the theoretical
prediction for its branching ratio, crucial role is played by the NLO
QCD corrections to the $b \to s \gamma$ partonic amplitude. In the
introduction to our previous article~\cite{Buras:2001mq}, status of the NLO
calculations has been summarized. The only missing elements were
two-loop matrix elements of the so-called QCD-penguin operators.

The present paper is devoted to evaluation of these matrix
elements. Our results for two-loop diagrams are presented in such a
manner that they can be applied to an arbitrary extension of the SM
where additional four-quark operators arise, e.g. to the generic
MSSM. Thus, apart from completing the NLO QCD calculation in the SM,
we provide an important ingredient for analyses of new physics
effects. Moreover, contrary to previous
calculations~\cite{Greub:1996tg,Buras:2001mq}, no expansion in the
mass ratio $m_c/m_b$ is performed.

The article is organized as follows. In the next section, the relevant
definitions are collected. In section~\ref{sec:res}, we summarize our
results for the matrix elements, and describe consequences for
BR$[\bar{B} \to X_s \gamma]$ in the SM. Section~\ref{sec:M3Q1}
contains details of the calculation for one set of the two-loop
diagrams. For the remaining sets, we present only the final
expressions in section~\ref{sec:expr}.  The renormalization procedure
is described in section~\ref{sec:renor}. Section~\ref{sec:concl}
contains our conclusions.

\newsection{The effective Hamiltonian}
\label{sec:hamil}

In the SM, the $b\to s \gamma$ transition is mediated by the effective
Hamiltonian\footnote{
It is written in terms of bare quantities here. The
renormalized interaction terms can be found in eq.~(\ref{lagr}).}${}^{,}\!\!$
\footnote{For simplicity, we set $V_{ub}$ to zero in all our formulae.
However, its non-zero value will be included in the phenomenological
analysis, as in eq.~(3.7) of ref.~\cite{Gambino:2001ew}.}
\be \label{hamil}
{\cal H}_{\mathrm{eff}} = -\f{4 G_F}{\sqrt{2}} V^*_{ts} V_{tb} \sum_{k=1}^8 C_k P_k,
\ee
where $C_k$ are the Wilson coefficients and $P_k$ stand for the
following operators:
\bea 
P_1  &=& (\bar{s}_L \gamma_{\mu} T^a c_L)        (\bar{c}_L \gamma^{\mu} T^a b_L),\nonumber\\
P_2  &=& (\bar{s}_L \gamma_{\mu}     c_L)        (\bar{c}_L \gamma^{\mu}     b_L),\nonumber\\
P_3  &=& (\bar{s}_L \gamma_{\mu}     b_L) \sum_q (\bar{q}   \gamma^{\mu}     q),  
\hspace{3cm} (q=u,d,s,c,b) \nonumber\\
P_4  &=& (\bar{s}_L \gamma_{\mu} T^a b_L) \sum_q (\bar{q}   \gamma^{\mu} T^a q),  \nonumber\\
P_5  &=& (\bar{s}_L \gamma_{\mu}
                    \gamma_{\nu}
                    \gamma_{\rho}    b_L) \sum_q (\bar{q}   \gamma^{\mu} 
                                                            \gamma^{\nu}
                                                            \gamma^{\rho}    q),  \nonumber\\
P_6  &=& (\bar{s}_L \gamma_{\mu}
                    \gamma_{\nu}
                    \gamma_{\rho}T^a b_L) \sum_q (\bar{q}   \gamma^{\mu} 
                                                            \gamma^{\nu}
                                                            \gamma^{\rho}T^a q),  \nonumber\\
P_7  &=&  \f{e}{16 \pi^2} m_b (\bar{s}_L \sigma^{\mu \nu}     b_R) F_{\mu \nu},
\nonumber\\[1mm]
P_8  &=&  \f{g}{16 \pi^2} m_b (\bar{s}_L \sigma^{\mu \nu} T^a b_R) G_{\mu \nu}^a.
\label{ops}
\eea
In order to match our computation with the NLO results for $C_k$, we
adopt here the same operator basis as in ref.~\cite{Chetyrkin:1997vx}.

Instead of the original Wilson coefficients $C_k$ it is convenient to
use certain linear combinations of them, the so-called ``effective
coefficients'' \cite{Buras:1994xp}
\be \label{ceff}
C_k^{\mathrm{eff}} = \left\{ \begin{array}{lcl}
C_k, &~~~& \mbox{ for $k = 1, ..., 6$}, \\ 
C_7 + \sum_{i=1}^6 y_i C_i, && \mbox{ for $k = 7$}, \\
C_8 + \sum_{i=1}^6 z_i C_i, && \mbox{ for $k = 8$}.
\end{array} \right.
\ee
The numbers $y_i$ and $z_i$ are defined so that the leading-order $b
\to s \gamma$ and $b \to s\;gluon$ matrix elements of the effective
Hamiltonian are proportional to the leading-order terms in $C_7^{\mathrm{eff}}$
and $C_8^{\mathrm{eff}}$, respectively. In the NDR scheme,
\be \label{yyzz}
\vec{y} = \left(0, 0,-\f{1}{3},-\f{4}{9},-\f{20}{3},-\f{80}{9}\right),
\hspace{2cm}
\vec{z} = \left(0, 0, 1,-\f{1}{6}, 20, -\f{10}{3}\right).
\ee
The renormalization group equations for the $\overline{\rm
MS}$-renormalized effective coefficients read
\be 
\mu \f{d}{d \mu} C_i^{\mathrm{eff}}(\mu) = 
C_j^{\mathrm{eff}}(\mu) \gamma^{\mathrm{eff}}_{ji}(\mu).
\ee
Explicit results for the coefficients 
\be
C^{\mathrm{eff}}_i(\mu) =
C_i^{(0)\mathrm{eff}}(\mu) + \f{\al(\mu)}{4\pi} C_i^{(1)\mathrm{eff}}(\mu) 
+ {\cal O}(\al^2)
\ee
and the matrix 
\be
\hat{\gamma}^{\mathrm{eff}} = \f{\al}{4\pi}
\hat{\gamma}^{(0)\mathrm{eff}} + \left(\f{\al}{4\pi}\right)^2
\hat{\gamma}^{(1)\mathrm{eff}} + {\cal O}(\al^3)
\ee
can be found in
ref.~\cite{Chetyrkin:1997vx} (see also eq.~(\ref{gamma0}) here). The partonic
decay rate from eqs. (29)--(32) of that article can be expressed as
follows:
\be
\Gamma[b\to X_s \gamma]^{E_{\gamma} > E_0} = \f{ G_F^2 \alpha_{em} m_b^5}{32 \pi^4} 
\left| V_{ts}^* V_{tb} \right|^2 \left[ |D|^2 + A + {\cal O}(\al^2,\alpha_{em}) \right],
\ee
where, for $\mu \sim m_b$,
\mathindent0cm
\bea \label{Dfactor}
D &=& \left({\rm Terms~proportional~to~} C_7^{\mathrm{eff}}(\mu) \right) \;\;+\;\; 
\f{\al}{4\pi} \sum_{\begin{array}{c}\\[-7mm] 
{\scs 1\leq k\leq 8}\\[-2mm] {\scs k\neq 7} \end{array}} 
C_k^{(0)\mathrm{eff}}(\mu) \left[ r_k +\gamma^{(0)\mathrm{eff}}_{k7}\ln\f{m_b}{\mu}\right],
\\[2mm]
A &=&      \left({\rm Terms~proportional~to~} |C_7^{(0)\mathrm{eff}}(\mu)|^2\right) 
 \;\;+\;\; \left({\rm Terms~vanishing~when~} E_0 \to \f{m_b}{2} \right).
\eea
\mathindent1cm

\newpage
\begin{figure}[t]
\begin{center}
\includegraphics[width=17cm,angle=0]{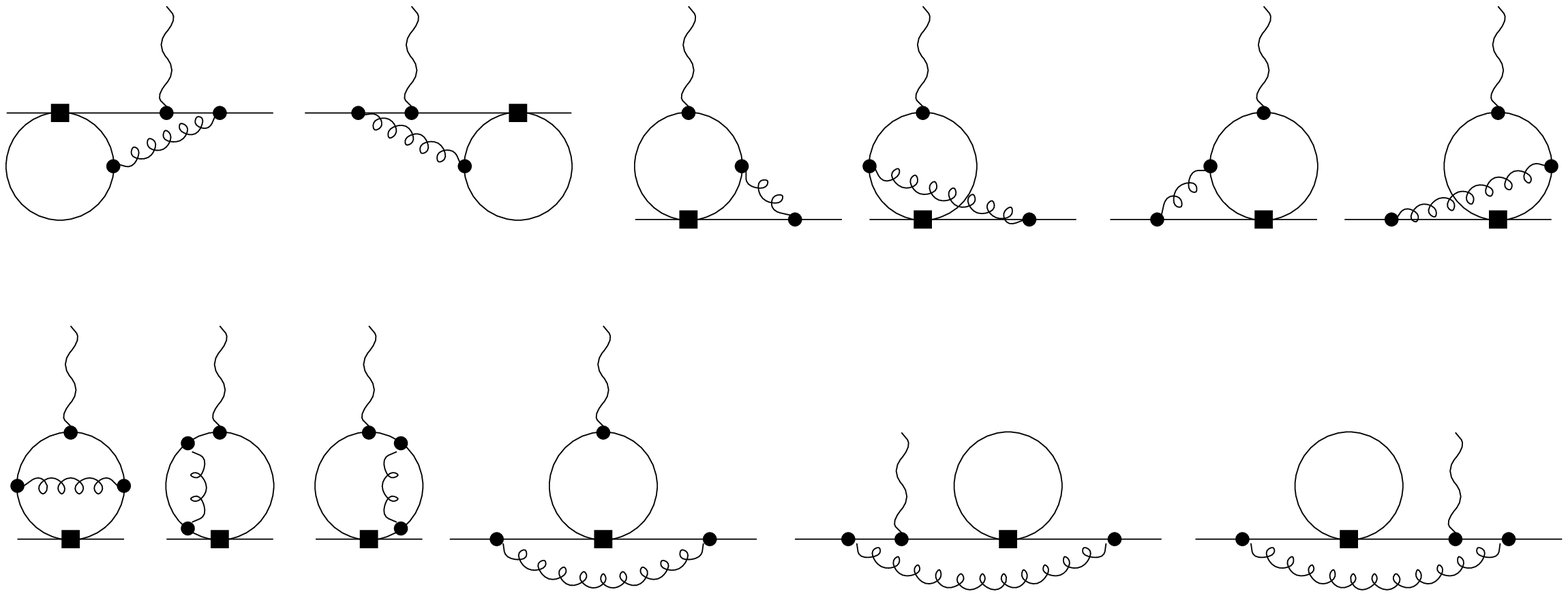}
\end{center}
\vspace*{-5cm}
\hspace*{12mm} $G_1$ \hspace{25mm} $G_2$\\[-3mm]
\hspace*{67mm} $\underbrace{\hspace{49mm}}~\underbrace{\hspace{49mm}}$\\[2mm]
\hspace*{9cm} $G_3$ \hspace*{44mm} $G_4$\\[22mm]
$\underbrace{\hspace{46mm}}$\\[-1mm]
\hspace*{85mm} $\underbrace{\hspace{83mm}}$\\[-2mm]
\hspace*{19mm} $G_5$ \hspace{35mm} $G_6$\\[-1mm]
\hspace*{125mm} $G_7$\\[-1cm] 
\begin{center}
\caption{\sf Two-loop 1PI contributions to the matrix elements of $P_k$. 
\label{fig:diagrams}}
\end{center}
\end{figure}
\ \\[-15mm] 
For $k=1,..,6$, the quantities $r_k$ in eq.~(\ref{Dfactor}) can be
found by calculating the two-loop $b\to s\gamma$ matrix elements of
the four-quark operators $P_k$. The relevant Feynman diagrams are
presented in fig.~\ref{fig:diagrams}, where the square boxes denote
the operator insertions. On the other hand, $r_8$ is given by the
one-loop matrix element of $P_8$.

\newsection{Final results for $r_k$ and consequences for 
BR$[\bar{B} \to X_s \gamma]$}
\label{sec:res}

Calculation of $r_k$ in eq.~(\ref{Dfactor}) has been the main goal of
the present paper. For $k=1,2,8$, we confirm the findings of
refs.~\cite{Greub:1996tg,Buras:2001mq}. For $k=3, ..., 6$, our results are
new. Altogether, they read
\be \label{rk}
\begin{array}{rcl}
r_1 &=& \f{833}{729} - \f{1}{3} [a(z)+b(z)] + \f{40}{243} i \pi,\\[2mm]
r_2 &=& -\f{1666}{243} + 2[a(z)+b(z)] - \f{80}{81} i \pi,\\[2mm]
r_3 &=& \f{2392}{243} + \f{8\pi}{3\sqrt{3}} + \f{32}{9} X_b - a(1) + 2 b(1) + \f{56}{81} i \pi,\\[2mm]
r_4 &=& -\f{761}{729} - \f{4\pi}{9\sqrt{3}} - \f{16}{27} X_b + \f{1}{6} a(1) + \f{5}{3} b(1) 
         + 2 b(z) - \f{148}{243} i \pi,\\[2mm]
r_5 &=& \f{56680}{243} + \f{32\pi}{3\sqrt{3}} + \f{128}{9} X_b - 16 a(1) + 32 b(1) 
         + \f{896}{81} i \pi,\\[2mm]
r_6 &=& \f{5710}{729} - \f{16\pi}{9\sqrt{3}} - \f{64}{27} X_b - \f{10}{3} a(1) + \f{44}{3} b(1) 
         + 12 a(z) + 20 b(z) - \f{2296}{243} i \pi,\\[2mm]
r_8 &=& \f{44}{9} - \f{8}{27} \pi^2 + \f{8}{9} i \pi,
\end{array}
\ee
where $z = m_c^2/m_b^2$. Similar ratios for the light quarks
($u$, $d$, $s$) have been set to zero. The constant $X_b$ is given by
\be
X_b = \int_0^1 dx \int_0^1 dy \int_0^1 dv \; x y 
\ln\left[v + x (1-x) (1-v) (1-v + v y)\right] \simeq -0.1684.
\ee
Exact expressions for the functions $a(z)$ and $b(z)$ are as follows:
\mathindent0cm
\bea
a(z) &=& \fm{8}{9} \int_0^1 dx \int_0^1 dy \int_0^1 dv \left\{  
[2 - v + x y (2 v-3)] \ln [v z + x (1 - x) (1 - v) (1 - v + v y)] 
\right. \nonumber\\[2mm] && \hspace{17mm} \left. 
+[1-v+xy(2v-1)] \ln [ z - i\ve - x(1-x)yv] \right\} + \fm{43}{9} + \fm{4}{9} i \pi,\\[4mm]
b(z) &=& \fm{4}{81}\ln z +\fm{16}{27}z^2 +\fm{224}{81}z -\fm{92}{243} +\fm{4}{81}i\pi 
+\fm{-48z^2-64z+4}{81}\sqrt{1\!-\!4z} f(z)
-\fm{8}{9} z^2 \left( \fm{2}{3}z-1 \right) f(z)^2
\nonumber\\[2mm] && 
- \fm{8}{9} \int_0^1 dx \int_0^1 dy\; \f{\fm{1}{2} y^2 (y^2-1) x (1-x) 
+ (2-y) u_1 \ln u_1 + (2y^2-2y-1) u_2 \ln u_2}{(1-y)^2}, \label{bzz}
\eea
\mathindent1cm
with ~$u_k = y^k x(1-x)+(1-y) z$,~~ 
$\sqrt{-1} = +i$
~ and
\be \label{fzz} 
f(z) = \theta(1\!-\!4z) \left( \ln \fm{1+\sqrt{1\!-\!4z}}{1-\sqrt{1\!-\!4z}} 
-i \pi \right) -2i \theta(4z\!-\!1) \arctan \fm{1}{\sqrt{4z\!-\!1}}.
\ee
Additive constants in the functions $a$ and $b$ have been chosen in
such a manner that $a(0)=b(0)=0$. Our results for $a(1)$ and
$b(1)$ read
\bea
a(1) &\simeq& 4.0859 + \f{4}{ 9} i \pi,\\
b(1) &=& \f{320}{81} - \f{4 \pi}{3 \sqrt{3}} 
+ \f{632}{1215} \pi^2 - \f{8}{45} 
\left[ \f{d^2 \ln \Gamma(x)}{dx^2} \right]_{x=\f{1}{6}}
+ \f{4}{81} i \pi
~~\simeq~~ 0.0316 + \f{4}{81} i \pi.
\eea
%
%
\begin{figure}[t]
\begin{center}
\vspace*{-3cm}
\includegraphics[width=8cm,angle=0]{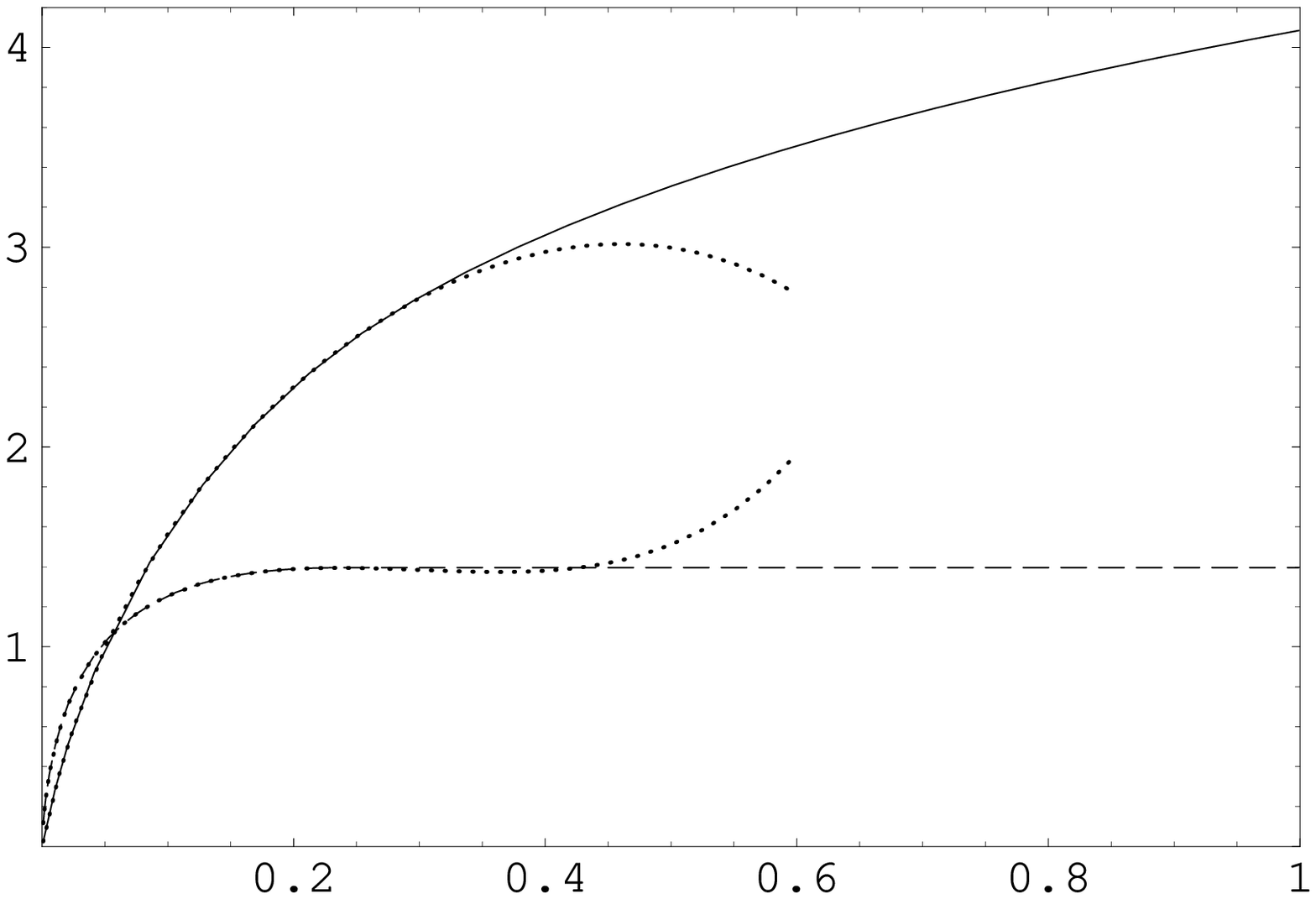}
\hspace{5mm}
\includegraphics[width=8cm,angle=0]{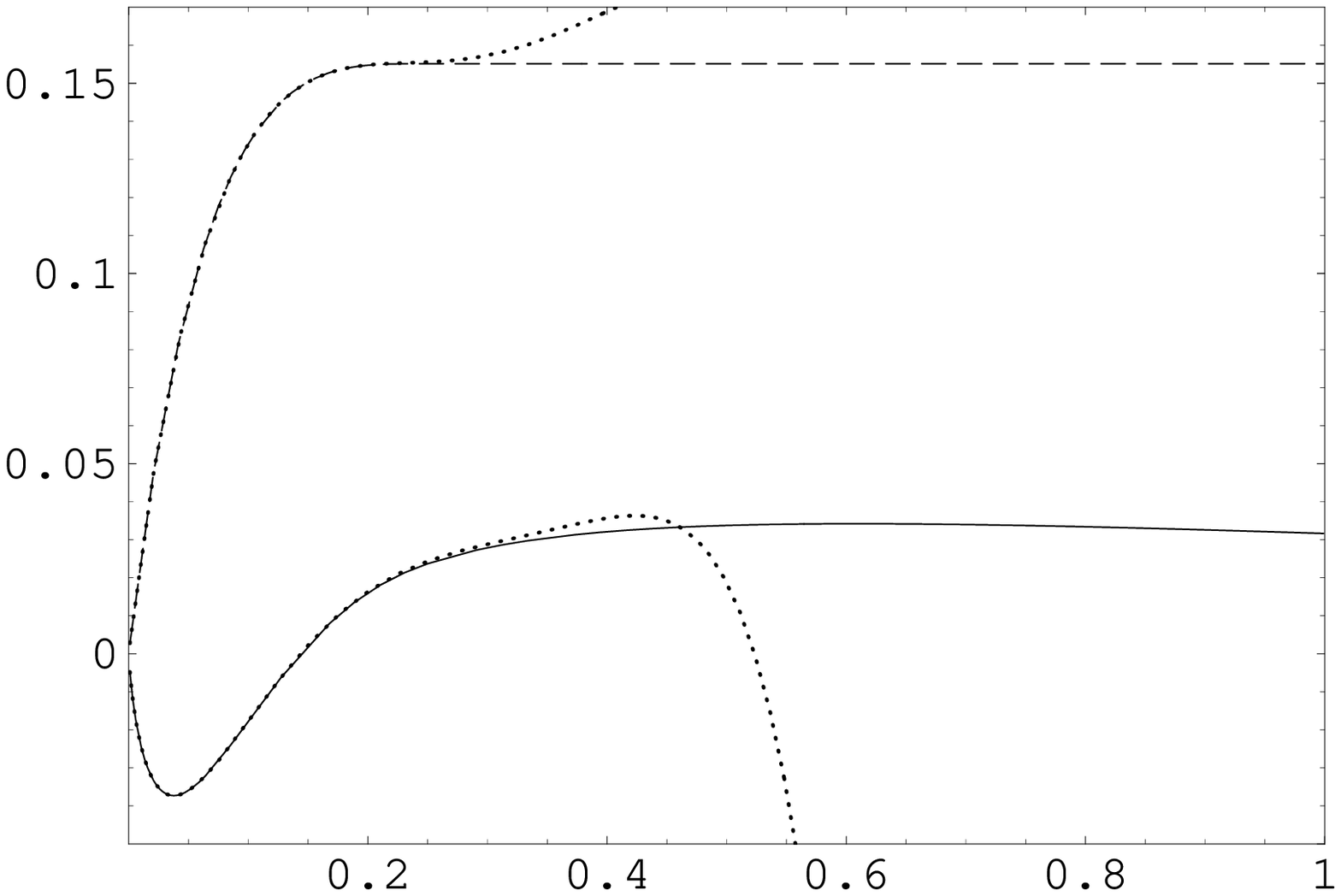}
\end{center}
\vspace*{-84mm}
\hspace*{62mm} Re $a(z)$  \hspace{72mm} Im $b(z)$\\[16.5mm]
\hspace*{151mm} Re $b(z)$\\[-3mm]
\hspace*{62mm} Im $a(z)$\\[12mm]
\hspace*{75mm} $z$  \hspace{83mm} $z$
\begin{center}
\caption{\sf Functions a(z) and b(z). Dotted lines represent their
expansions at $z=0$ up to ${\cal O}(z^6)$. \label{fig:ab}}
\end{center}
\vspace*{-1cm}
\end{figure}
There is no need to apply numerical integration for
$z=m_c^2/m_b^2 \sim 0.1$, because, as illustrated in
fig.~\ref{fig:ab}, both functions are then accurately given by their
expansions in $z$~~\cite{Greub:1996tg,Buras:2001mq},
\mathindent0cm
\bea \ \nonumber\\[25mm] \ \\[-37mm] 
\begin{array}{rcl}
a(z)  &=& \f{16}{9} \left\{ 
\left[ \f{5}{2} -\f{1}{3}\pi^2 -3 \zeta(3) + \left( \f{5}{2} - \f{3}{4} \pi^2 \right) L
+ \f{1}{4} L^2 + \f{1}{12} L^3 \right] z 
\right. \\[2mm] && \hspace{3mm} \left.
+\left[\f{7}{4} +\f{2}{3} \pi^2 -\f{1}{2} \pi^2 L -\f{1}{4} L^2 +\f{1}{12} L^3 \right] z^2 
+ \left[ -\f{7}{6} -\f{1}{4} \pi^2 + 2L - \f{3}{4}L^2 \right] z^3
\right. \\[2mm] && \hspace{3mm} \left.
+\left[ \f{457}{216} - \f{5}{18} \pi^2 -\f{1}{72} L -\f{5}{6} L^2 \right] z^4 
+\left[ \f{35101}{8640} - \f{35}{72} \pi^2 -\f{185}{144} L -\f{35}{24} L^2 \right] z^5
\right. \\[2mm] && \hspace{3mm} \left.
+ \left[ \f{67801}{8000} - \f{21}{20} \pi^2 - \f{3303}{800} L - \f{63}{20} L^2 \right] z^6 
+ i \pi \left[ \left( 2 -\f{1}{6} \pi^2 + \f{1}{2} L + \f{1}{2} L^2 \right) z 
\right. \right. \\[2mm] && \hspace{3mm} \left. \left.
+ \left( \f{1}{2} -\f{1}{6} \pi^2 - L + \f{1}{2} L^2 \right) z^2 + z^3 
+ \f{5}{9} z^4 + \f{49}{72} z^5  + \f{231}{200} z^6 \right] \right\} + {\cal O}(z^7 L^2),
\end{array} \nonumber \eea 
\bea \begin{array}{rcl}
b(z) &=& \hspace{-2mm} -\frac{8}{9} \left\{
\left( -3 +\f{1}{6} \pi^2 - L \right) z - \f{2}{3}\pi^2 z^{3/2}
+ \left( \f{1}{2} + \pi^2 - 2 L - \f{1}{2} L^2 \right) z^2 
\right. \\[2mm] && \hspace{3mm} \left. 
+ \left(-\f{25}{12} -\f{1}{9} \pi^2 - \f{19}{18} L + 2 L^2 \right) z^3 
+ \left[ -\f{1376}{225} + \f{137}{30} L + 2 L^2 + \f{2}{3} \pi^2 \right] z^4 
\right. \\[2mm] && \hspace{3mm} \left. 
+ \left[ -\f{131317}{11760} + \f{887}{84} L + 5 L^2 + \f{5}{3} \pi^2 \right] z^5 
+ \left[ -\f{2807617}{97200} + \f{16597}{540} L + 14 L^2 + \f{14}{3} \pi^2 \right] z^6 
\right. \\[2mm] && \hspace{3mm} \left. 
+ i \pi \left[ -z + (1 - 2 L) z^2 + \left(-\f{10}{9} + \f{4}{3} L \right) z^3 
      + z^4 + \f{2}{3} z^5 + \f{7}{9} z^6 \right] \right\} + {\cal O}(z^7 L^2),
\end{array} \nonumber\\[-6.5mm] \eea
\mathindent1cm
\ \\[-5mm]
where $L = \ln z$.\\

\begin{table}[h]
\begin{center}
\begin{tabular}{|r|r|r||r|r|}
\hline
      & \multicolumn{2}{|c||}{\TT$\f{m_c}{m_b}=0.22$} &
        \multicolumn{2}{|c|}{$\f{m_c}{m_b}=0.29$}\\[2mm]
\hline
 &\TT Re $r_i$~~  & Im $r_i$~~           & Re $r_i$~~  & Im $r_i$~~ \\[2mm]
\hline
$r_1$ & $0.8309$  & $0.1498$             & $0.6821$  & $0.0750$   \\[2mm]
$r_2$ & $-4.9854$ & $-0.8988$            & $-4.0929$ & $-0.4499$  \\[2mm]
$r_3$ & $10.0589$ & $1.0860$             & $10.0589$ & $1.0860$   \\[2mm]
$r_4$ & $-1.0890$ & $-1.2409$            & $-1.0655$ & $-1.1732$  \\[2mm]
$r_5$ & $185.8412$& $17.3757$            & $185.8412$& $17.3757$  \\[2mm]
$r_6$ & $2.7855$  & $-18.1132$           & $8.2345$  & $-15.1493$ \\[2mm]
$r_8$ & $1.9646$  & $2.7925$             & $1.9646$  & $2.7925$   \\
\hline
\end{tabular}
\end{center}
\caption{\sf Real and imaginary parts of $r_k$ for two different values of
$m_c/m_b$. \label{tab:rk}}
\end{table}
The numerical values of $r_k$ for two different values of $m_c/m_b$
are presented in table~\ref{tab:rk}.  We observe that the real parts
of $r_3$ and $r_5$ are considerably larger than the remaining ones.
If $r_4$ were as large as $r_3$, its effect on the SM branching
ratio would be around 6\%. On the other hand, the size of $r_5$ is
somewhat artificial --- it is due in part to the lack of a ``natural''
factor $\fm{1}{16}$ in the definition of $P_5$ (\ref{ops}).

In order to calculate BR$[\bar{B} \to X_s \gamma]$, we use the
NLO formulae collected in ref.~\cite{Gambino:2001ew}. The only elements that
must be modified due to non-zero values of $r_3$, ..., $r_6$ are the
so-called ``magic numbers''. Their updated values are presented in
table~\ref{tab:mag}. Once they are used, and the same computer program
with the same numerical inputs is applied,\footnote{
One exception is $\varepsilon_{ew}$ from eq.~(4.6) of
ref.~\cite{Gambino:2001ew} which we change to 0.0071 according to
ref.~\cite{Gambino:2001au}. This modification slightly diminishes the net effect
on the branching ratio.}
we find
\be 
{\rm BR}[\bar{B} \to X_s \gamma]^{{\rm subtracted~} \psi,\;\psi'
}_{E_{\gamma} > 1.6~{\rm GeV}}
= (3.57 \pm 0.30) \times 10^{-4},
\label{main.num}
\ee
which is only a little lower than $(3.60 \pm 0.30) \times 10^{-4}$ in
eq.~(4.14) of ref.~\cite{Gambino:2001ew}. For the cutoff energy of $\f{1}{20}
m_b$ (which corresponds to the fake ``total rate''), the effect on the
branching ratio is the same: the central value decreases from
$3.73\times 10^{-4}$ to $3.70 \times 10^{-4}$.

Such a small effect of the non-zero values of $r_3$, ..., $r_6$ is due
to smallness of the Wilson coefficients of the corresponding
operators. In ref.~\cite{Chetyrkin:1997vx}, this effect was estimated to be
around 1\%. For that reason, the NLO QCD computation was called
``practically complete'' already at that time.  However, it becomes
{\em strictly complete} only now, when we learn exactly what the
effect is and in what direction it acts.

\begin{table}[t]
\begin{tabular}{|l|r|r|r|r|r|r|r|r|}
\hline
~~~$k$ & 1 & 2 & 3 & 4 & 5~~~~ & 6~~~~ & 7~~~~ & 8~~~~ \\ 
\hline
\ &&&&&&&& \\[-4mm]
$a_k$              & $\f{14}{23}$~~~  &  $\f{16}{23}$~~~ & $\f{6}{23}$~~~ & $-\f{12}{23}$~~~ 
                   &     0.4086 & $-$0.4230 & $-$0.8994 &    0.1456 \\[1.5mm]
$d_k$              &     1.4107 & $-$0.8380 & $-$0.4286 & $-$0.0714 
                   &  $-$0.6494 & $-$0.0380 & $-$0.0185 & $-$0.0057 \\[1.5mm]
$\tilde{d}_k$      & $-$17.6507 &   11.3460 &    2.4692 & $-$0.8056
                   &     4.8898 & $-$0.2308 & $-$0.5290 &    0.1994\\[1.5mm]
$\tilde{d}^{\eta}_k$&    9.2746 & $-$6.9366 & $-$0.8740 &    0.4218 
                   &  $-$2.7231 &    0.4083 &    0.1465 &    0.0205\\[1.5mm]
$\tilde{d}^{a}_k$  &   0~~~~~~~~&  0~~~~~~~~&    0.8571 &    0.6667 
                   &     0.1298 &    0.1951 &    0.1236 &    0.0276\\[1.5mm]
$\tilde{d}^{b}_k$  &   0~~~~~~~~&  0~~~~~~~~&    0.8571 &    0.6667 
                   &     0.2637 &    0.2906 & $-$0.0611 & $-$0.0171\\[1.5mm]
$ \tilde{d}^{i \pi}_k$&  0.4702 &  0~~~~~~~~& $-$0.4268 & $-$0.2222 
                   &  $-$0.9042 & $-$0.1150 & $-$0.0975 &    0.0115\\[1.5mm]
$ e_k$             &     5.2620 & $-$3.8412 &  0~~~~~~~~& 0~~~~~~~
                   &  $-$1.9043 & $-$0.1008 &    0.1216 &    0.0183 \\[1.5mm]
\hline
\end{tabular}
\caption{\sf Update of table 1 from ref.~\cite{Gambino:2001ew}. Only
$\tilde{d}_k$, $\tilde{d}^{a}_k$, $\tilde{d}^{b}_k$, $ \tilde{d}^{i
\pi}_k$ for $k=3, ..., 8$ are affected. \label{tab:mag}}
\end{table}

\newsection{Example of a two-loop diagram calculation: $G_3$}
\label{sec:M3Q1}

In the present section, we describe in detail the calculation of the
two diagrams denoted by $G_3$ \linebreak in
fig.~\ref{fig:diagrams}. We choose $(\bar{s}_L \gamma_{\mu}
c_L)(\bar{c}_L \gamma^{\mu} b_L)$ for the inserted operator. Here,
contrary to refs.~\cite{Buras:2001mq,Greub:1996tg}, \linebreak no
expansion in $z$ is performed, because the limit $z \to 1$ is relevant
for operators that contain three $b$-quarks.  The UV divergences are
regularized dimensionally. We treat $\gamma_5$ as fully anticommuting
in $D=4-2\e$ dimensions. The $s$-quark is assumed to be massless.

The considered Feynman integral can be written as follows
\be \label{G3.b}
G_3^{(1)} = \int \f{d^Dk}{k^2(q+k)^2}
\bar{s} \gamma^{\nu} ( \slash q + \slash k ) 
\gamma_{\rho} P_L J_{\mu \nu} e^{\mu} \gamma^{\rho} P_L b,
\ee
where 
\mathindent0cm
\be
J_{\mu \nu} = \int \f{d^D p}{\Delta} \left\{
[\slash p \!+\! \slash k \!+\! m_c] \gamma_{\nu} 
[\slash p \!+\! m_c] \gamma_{\mu} 
[\slash p \!+\! \slash r \!+\! m_c] 
\;-\;
[\slash p \!+\! \slash r \!-\! m_c] \gamma_{\mu} 
[\slash p \!-\! m_c] \gamma_{\nu} 
[\slash p \!+\! \slash k \!-\! m_c] \right\}
\label{JM3}
\ee
\mathindent1cm
and
\be
\Delta = [(p+k)^2-m_c^2][(p+r)^2-m_c^2][p^2-m_c^2].
\ee
Here, $r$ and $q$ stand for the outgoing photon and $s$-quark momenta,
respectively. The momentum $p$ runs inside the $c$-quark loop, while
$k$ is the gluon momentum. The polarization vector of the photon is
denoted by $e^{\mu}$.

Let us first consider the one-loop subgraph $J_{\mu \nu}$ for
arbitrary off-shell momenta of the quarks, the gluon and the
photon. The necessary one-loop integrals read
\bea
\int d^D p/\Delta &\equiv& a,\\[2mm]
\int d^D p\; p_\alpha/\Delta &=& b k_\alpha + c r_\alpha, \label{bdef}\\[2mm]
\int d^D p\; p_\alpha p_\beta/\Delta &=& d m_c^2 g_{\alpha \beta} + e k_\alpha k_\beta
         + f r_\alpha r_\beta + g ( k_\alpha r_\beta + r_\alpha k_\beta). \label{gdef}
\eea
There is no need to consider ~$p_\alpha p_\beta p_\gamma$~ in the numerator, because
~$\slash p \gamma_\nu \slash p \gamma_\mu \slash p -
 \slash p \gamma_\mu \slash p \gamma_\nu \slash p =
p^2 ( \slash p \gamma_\mu \gamma_\nu - \gamma_\nu \gamma_\mu \slash p)$.~
The coefficients $a, ..., g$ are not all independent. The following
relations can be found by considering various contractions of the
tensor integrals:
\bea
b &=& -\f{1}{2}a - c - 2g - (c+2f) \f{k\cd r}{k^2},\\[2mm]
d &=& \f{1}{8\e m_c^2} \left[ a(k^2-4m_c^2) + 2c(k\cd r+k^2-3r^2) 
      + 4f( k\cd r - 2r^2) + 4g(k^2-2k\cd r) \right],\\[2mm]
e &=& \f{1}{4}a + \f{1}{2}c + g + (c+2f) \f{k\cd r+r^2}{2k^2}.
\eea
After performing the Dirac algebra with the above relations taken into
account, one arrives at the following result:
\be \label{Jmunu}
J_{\mu\nu} = b V^b_{\mu\nu} + c V^{c}_{\mu\nu} + f V^{f}_{\mu\nu} + g V^g_{\mu\nu}
+ m_c \left[ a P^a_{\mu\nu} + (c+2f) P^{cf}_{\mu\nu} + g P^g_{\mu\nu} \right],
\ee
where
\bea
V^b_{\mu\nu} &=& 2 \left( X^{(2)}_{\mu\nu} - X^{(3)}_{\mu\nu} \right), \nnb\\ 
V^c_{\mu\nu} &=& 4 \left( \wi{X}^{(3)}_{\mu\nu} - \wi{X}^{(2)}_{\mu\nu} \right)
               + \f{2r^2}{k^2} \left( X^{(2)}_{\mu\nu} - X^{(3)}_{\mu\nu} \right),\nnb\\ 
V^f_{\mu\nu} &=& 4 \left( \wi{X}^{(3)}_{\mu\nu} - \wi{X}^{(2)}_{\mu\nu} \right)
               + \f{4r^2}{k^2} \left( X^{(2)}_{\mu\nu} - X^{(3)}_{\mu\nu} \right),\nnb\\ 
V^g_{\mu\nu} &=& 4 \left[ (\slash r - \slash k) X^{(1)}_{\mu\nu} 
                        + X^{(4)}_{\mu\nu} - \wi{X}^{(4)}_{\mu\nu} \right], \nnb\\
P^a_{\mu\nu} &=& 2 \left( X^{(5)}_{\mu\nu} - X^{(6)}_{\mu\nu} \right), \nnb\\
P^{cf}_{\mu\nu} &=& -\f{4}{k^2} X^{(7)}_{\mu\nu}, \nnb\\
P^{g}_{\mu\nu} &=& -8 X^{(1)}_{\mu\nu}, 
\eea
and
\bea
X^{(1)}_{\mu\nu} &=& k \cd r g_{\mu\nu} - k_\mu r_\nu,\nnb\\
X^{(2)}_{\mu\nu} &=&   \slash r (k^2 g_{\mu\nu} - k_\mu k_\nu) 
                     - \gamma_\mu ( k^2 r_\nu - k\cd r\, k_\nu),\nnb\\
X^{(3)}_{\mu\nu} &=& i (k^2 \gamma_\nu - \slash k k_\nu) \sigma_{\alpha \mu} r^\alpha,\nnb\\
X^{(4)}_{\mu\nu} &=& i (k\cd r \gamma_\mu - \slash r k_\mu) \sigma_{\alpha \nu} k^\alpha,\nnb\\
X^{(5)}_{\mu\nu} &=& \sigma_{\alpha\nu} k^\alpha \; \sigma_{\beta\mu} r^\beta,\nnb\\
X^{(6)}_{\mu\nu} &=& i \left[   \sigma_{\mu\nu} k\cd r 
                              - \sigma_{\alpha\beta} k^\alpha r^\beta g_{\mu\nu}
                              + \sigma_{\alpha\mu} k^\alpha r_\nu
                              - \sigma_{\alpha\nu} r^\alpha k_\mu \right],\nnb\\
X^{(7)}_{\mu\nu} &=& k^2 r^2 g_{\mu\nu} - r^2 k_\mu k_\nu - k^2 r_\mu r_\nu 
                                                       + k \cd r k_\nu r_\mu.
\eea
The structures ~$\wi{X}^{(n)}_{\mu\nu}$~ are obtained from
~$X^{(n)}_{\mu\nu}$~ by interchanging ~$\mu \leftrightarrow \nu$~ and
~$k \leftrightarrow r$,~ i.e.\linebreak
$\wi{X}^{(n)}_{\mu\nu}(k,r) = X^{(n)}_{\nu\mu}(r,k)$.  Note
that each of those structures vanishes under contractions with $k^\nu$
and $r^\mu$.  This is a manifestation of the Ward identities $k^\nu
J_{\mu\nu} = r^\mu J_{\mu\nu} =0$.

Once we have found the one-loop integral $J_{\mu\nu}$, we should
complete the remaining Dirac algebra in $G_3^{(1)}$ (\ref{G3.b}). 
Given our explicit expressions for the structures $X^{(k)}_{\mu\nu}$
and $\wi{X}^{(k)}_{\mu\nu}$, it is easy to verify that
\be \label{gJg}
\gamma_\rho P_L J_{\mu\nu} \gamma^{\rho} P_L =
(2+2\e) P_R \left[ b V^b_{\mu\nu} + c V^{c}_{\mu\nu} + f V^{f}_{\mu\nu} + g V^g_{\mu\nu} \right].
\ee
Consequently, in our present calculation of the matrix element of
$(\bar{s}_L \gamma_{\mu} c_L)(\bar{c}_L \gamma^{\mu} b_L)$, the terms
proportional to $m_c$ in $J_{\mu\nu}$ (\ref{Jmunu}) are irrelevant.
Nevertheless, we have presented them explicitly here because they
matter for other operators.
 
From now on, we shall impose the on-shell conditions for the quarks
and the photon:\linebreak $q^2=r^2=0$,~ ~$q\cd r = \f{1}{2} m_b^2$,~
~$\bar s \slash q = 0$~ and ~$(\slash q + \slash r)b = m_b b$. For an
arbitrary scalar function $F(k^2,k\cd r, k\cd q)$, we find the
following identities:
\mathindent0cm
\bea
\int d^D k \; F(k^2, k\cd r, k\cd q) \bar s \gamma^\nu (\slash q + \slash k) V^b_{\mu\nu} e^\mu b &=&
[{\rm terms~proportional~to~} \bar s \slash e b {\rm ~and~} (e\cd r)\bar s b]
\nnb\\&&\hspace{-5cm}
+ m_b \bar s (\slash e \slash r - \slash r \slash e) b \int d^D k \; F(k^2, k\cd r, k\cd q)
\left[ (k+q)^2 \f{k\cd r}{m_b^2} - k^2 \left( 1 + \f{2 k\cd r}{m_b^2} \right) \right],\\[2mm]
\int d^D k \; F(k^2, k\cd r, k\cd q) \bar s \gamma^\nu (\slash q + \slash k) V^{c}_{\mu\nu} e^\mu b &=&
     [{\rm terms~proportional~to~} \bar s \slash e b {\rm ~and~} (e\cd r)\bar s b],\\[2mm]
\int d^D k \; F(k^2, k\cd r, k\cd q) \bar s \gamma^\nu (\slash q + \slash k) V^{f}_{\mu\nu} e^\mu b &=&
     [{\rm terms~proportional~to~} \bar s \slash e b {\rm ~and~} (e\cd r)\bar s b],\\[2mm]
\int d^D k \; F(k^2, k\cd r, k\cd q) \bar s \gamma^\nu (\slash q + \slash k) V^{g}_{\mu\nu} e^\mu b &=&
[{\rm terms~proportional~to~} \bar s \slash e b {\rm ~and~} (e\cd r)\bar s b] 
\nnb \\[2mm] && \hspace{-77mm}
+\f{m_b \bar s (\slash e \slash r - \slash r \slash e) b}{1-\e} 
\int d^D k \; F(k^2, k\cd r, k\cd q) \left[ 2\e (k+q)^2 \f{k\cd r}{m_b^2} 
- k^2 \left( 1+(2\! -\! 3\e\! +\! 2\e^2) \f{2k\cd r}{m_b^2} 
\right) \right].
\eea
\mathindent1cm
Consequently, $G_3^{(1)}$ (\ref{G3.b}) takes the following form
\be \label{G3.1}
G_3^{(1)} = 4\Gamma(1\!+\! 2\e) \pi^D m_b^{-4\e} ~~ \bar{s} P_R \left\{ 
  R_3^{(1)} m_b (\slash e \slash r - \slash r \slash e) 
+ R^{'(1)}_3 m_b\;e\cd r + R^{''(1)}_3 \slash e~ \right\} b,
\ee
where
\bea
R_3^{(1)} &=& \f{(1\! +\! \e)m_b^{4\e}}{2\Gamma(1\! +\! 2\e) \pi^D} 
\int \f{d^Dk}{k^2(k+q)^2} \left\{ b(k^2,k\cd r) \left[ (k+q)^2 \f{k\cd r}{m_b^2} 
-k^2 \left( 1 + \f{2 k\cd r}{m_b^2} \right) \right]
\right. \nnb \\[2mm] && \left. \hspace{2cm}
+ \f{g(k^2,k\cd r)}{1-\e} \left[ 2\e (k+q)^2 \f{k\cd r}{m_b^2} 
- k^2 \left( 1+(2\! -\! 3\e\! +\! 2\e^2) \f{2k\cd r}{m_b^2} \right) \right] \right\}. 
\label{R3.1}
\eea
As discussed in the next section, we do not need to calculate
$R^{'(1)}_3$ and $R^{''(1)}_3$.  The overall normalization factor in
eq.~(\ref{G3.1}) is chosen in such a manner that the ${\cal O}(1)$
part of $R_3^{(1)}$ is equal to the quantity $\wi{M}_3$ from our
previous article~\cite{Buras:2001mq} (see eqs.~(2.6) and (2.11) there).

The terms proportional to ~$(k+q)^2 k\cdot r$~ in the curly bracket of
eq.~(\ref{R3.1}) give vanishing integrals over $k$. Consequently,
$R_3^{(1)}$ simplifies to
\mathindent0cm
\be
R_3^{(1)} = -\f{(1\! +\! \e)m_b^{4\e}}{2\Gamma(1\! +\! 2\e) \pi^D} 
\int \f{d^Dk}{(k+q)^2+i\ve} \left[ \left( 1 + \f{2 k\cd r}{m_b^2} \right) b
+ \left( 1+(2\! -\! 3\e\! +\! 2\e^2) \f{2k\cd r}{m_b^2} \right) \f{g}{1-\e} \right].
\ee
\mathindent1cm
Now, we need explicit expressions for $b$ and $g$. They can be
easily found from eqs.~(\ref{bdef}) and (\ref{gdef}) with the help of
Feynman parameters
\bea
\label{deffuncb}
b(k^2,k\cd r) &=& i \pi^{D/2} \Gamma(1+\e) \int_0^1 dx \int_0^1 dy \;
\f{x^{-\e} (1-x)^{-\e}}{[ -(k-yr)^2 + \f{m_c^2-i\ve}{x(1-x)}]^{1+\e}},\\[2mm]
\label{deffuncg}
g(k^2,k\cd r) &=& i \pi^{D/2} \Gamma(1+\e) \int_0^1 dx \int_0^1 dy \;
\f{x^{-\e} (1-x)^{-\e}}{[ -(k-yr)^2 + \f{m_c^2-i\ve}{x(1-x)}]^{1+\e}} (-xy).
\eea 
Introducing another Feynman parameter and integrating over $k$, one
obtains
\mathindent0cm
\be \label{R3e}
R_3^{(1)} = -\f{1+\e}{4\e(1-\e)} \int_0^1 dx \int_0^1 dy \int_0^1 dv \;
\f{ (1-\e) v + xy [ (1-v) (2-3\e+2\e^2) -1 ] }{
x^{-\e} (1-x)^{-\e} \, v^\e \, [ z -i\ve - x(1-x)y (1-v) ]^{2\e}},
\ee
\mathindent1cm
where $z = m_c^2/m_b^2$. 
%
%
After expanding the integrand to ${\cal O}(\e)$ and performing the
easy integrations, we find
\mathindent0cm
\be 
R_3^{(1)} = -\f{1}{8\e} + \f{1}{8} + \f{1}{2} \int_0^1 dx \int_0^1 dy \int_0^1 dv \;
[1-v+xy(2v-1)] \ln [ z - i\ve - x(1-x)yv].
\ee
\mathindent1cm
In the next section, analogous results for all the diagrams in
fig.~\ref{fig:diagrams} will be presented.

\newsection{Results for the unrenormalized two-loop diagrams}
\label{sec:expr}

In the present section, we shall be interested in the {\em
unrenormalized} two-loop 1PI on-shell \linebreak $b \to s \gamma$
diagrams with insertions of the following 4-quark operators:
\bea
Q^{(n)}  &=& 
\f{1}{4^{n-1}} 
               (\bar s P_R \gamma_{\alpha_1} ... \gamma_{\alpha_n} c)
               (\bar c     \gamma^{\alpha_n} ... \gamma^{\alpha_1} P_L b),\\
\wi{Q}^{(n)} &=& \f{1}{4^{n-1}} 
               (\bar s P_R \gamma_{\alpha_1} ... \gamma_{\alpha_n} c)
               (\bar c     \gamma^{\alpha_n} ... \gamma^{\alpha_1} P_R b).
\eea
The corresponding diagrams in fig.~\ref{fig:diagrams} will be denoted
by $G_k^{(n)}$ and $\wi{G}_k^{(n)}$, respectively. Their overall
normalization is assumed to be the same as in eq.~(\ref{G3.b}), e.g.
\be
\wi G_3^{(n)} = \f{1}{4^{n-1}} 
\int \f{d^Dk}{k^2(q+k)^2}
\bar{s} \gamma^{\nu} ( \slash q + \slash k ) 
P_R \gamma_{\alpha_1} \ldots \gamma_{\alpha_n} J_{\mu \nu} e^{\mu} 
\gamma^{\alpha_n} \ldots \gamma^{\alpha_1} P_R b.
\ee

The operators $Q^{(0)}$, $Q^{(1)}$, $\wi{Q}^{(0)}$, $\wi{Q}^{(1)}$,
$\wi{Q}^{(2)}$ together with their mirror copies and analogous
operators with color-octet currents form a complete set of
dimension-six $(\bar s c)(\bar c b)$ operators in four spacetime
dimensions. Extending our results to $\Delta B = -\Delta S = 1$
four-quark operators with other flavor contents is straightforward,
as we perform no expansion in the ratio $m_c/m_b$. 

For the operators $Q^{(n)}$, only the diagrams $G_1^{(n)}$, ...,
$G_4^{(n)}$ matter. The remaining ones in fig.~\ref{fig:diagrams}
vanish due to chirality conservation in the charm-quark loop and QED
gauge invariance.

For $\wi{Q}^{(n)}$, the set $\wi{G}_7^{(n)}$ gives no contribution
on-shell, i.e. it cancels with the corresponding 1PR diagrams,
because it is proportional to the matrix element of a two-quark
operator ~$\bar s P_R b$~ that vanishes by the equations of motion, up
to a total derivative
\be \label{EOM}
\bar s P_R b 
= -\f{1}{m_b} \left[ \bar s P_R ( \slash D - m_b ) b + 
 \bar s \bbuildrel{\slash D}_{}^{\leftarrow} P_L b -
\partial^{\mu} \left( \bar s \gamma_{\mu} P_L b \right) \right].
\ee

The diagram $\wi{G}_6^{(n)}$ is just a product of two one-loop
diagrams. It contains an IR divergence that cancels out only after
adding the corresponding bremsstrahlung corrections. However, for all
the $\wi{Q}^{(n)}$ except $\wi{Q}^{(0)}$ (which is of no interest in
the SM), the effect of those diagrams can be taken into account by the
standard replacement $C_7 \to C_7^{\mathrm{eff}}$ (see
section~\ref{sec:renor}). Thus, we shall ignore $\wi{G}_6^{(n)}$ in
this section, i.e. we shall consider only $\wi{G}_1^{(n)}$, ....,
$\wi{G}_5^{(n)}$ for the operators $\wi{Q}^{(n)}$.

By analogy to eq.~(\ref{G3.1}), we write
\be \label{Gk.n}
G_k^{(n)} = 4\Gamma(1+2\e) \pi^D m_b^{-4\e} ~~ \bar{s} P_R \left\{ 
  R_k^{(n)} m_b (\slash e \slash r - \slash r \slash e) 
+ R^{'(n)}_k m_b\;e\cd r + R^{''(n)}_k \slash e~ \right\} b,
\ee
and similarly for $\wi{G}_k^{(n)}$. Below, we shall present our
results for the coefficients in front of $(\slash e \slash r - \slash
r \slash e)$ only.  Due to QED gauge invariance, the remaining Dirac
structures must cancel out when all the 1PI and 1PR diagrams are
included.  No explicit calculation of the 1PR diagrams is necessary,
as they can give no $(\slash e \slash r - \slash r \slash e)$
structure.

We have found the following expressions for $R_k^{(n)}$ and
$\wi{R}_k^{(n)}$:
\mathindent0cm
\be \label{M1.final}
R_1^{(1)} = \f{1}{36\e} -\! \f{1}{18}\ln z - \! \f{2}{3}z^2 - \! \f{47}{18}z 
+ \! \f{37}{216} + \! \f{12z^2 \! + \! 16z \! - \! 1}{18}
\sqrt{1\!-\!4z} f(z) +\! \f{z^2(2z \! - \! 3)}{3}  f(z)^2, 
\ee
\bea
R_2^{(1)} = -\f{5}{36\e} +\! \f{35}{72} -\! \f{1}{2}z +\! 
\int_0^1 \! dx \! \int_0^1 \! dy\; \f{\f{1}{2} y^2 (y^2 \! - \! 1) 
x (1 \! - \! x) + (2 \! - \! y) u_1 \ln u_1 
+ (2y^2 \! - \! 2y \! - \! 1) u_2 \ln u_2}{(1-y)^2}, \nnb\\[-1cm] \nnb
\eea
\be \ \label{M2.final} \ee
\be \label{M3.final}
R_3^{(1)} = -\f{1}{8\e} + \! \f{1}{8} + \! \f{1}{2} \int_0^1 \! dx \! 
\int_0^1 \! dy \! \int_0^1 \! dv \;\;
[1-v+xy(2v-1)] \; \ln [ z - i\ve - x(1-x)vy],
\ee
\ \\[-7mm]
\be \label{M4.final}
R_4^{(1)} \! = \! -\f{1}{4\e} \! + \! \f{3}{4} \! + \! \f{1}{2} \int_0^1 \! dx 
\! \int_0^1 \! dy \! \int_0^1 \! dv \; [2 \! -\! v\! + \! x y (2 v \! - \! 3)] 
\ln [v z \! + \! x (1 \! - \! x) (1 \! - \! v) (1 \! - \! v \! + \! v y)],
\ee
\ \\[-7mm]
\be \label{Mt1.final}
\wi{R}_1^{(1)} =
\sqrt{z} \left\{ -\fm{1}{4\e} \!-\! \fm{11+4i\pi}{8} 
+ \e \left[ \fm{10\pi^2-85-44i\pi}{16} 
+ \!\! \int_0^1 \! dx \! \int_0^1 \! dy \! \int_0^1 \! dv \; \fm{v}{y} \ln 
\left( 1 \!-\! \fm{yz}{x(1-x)(1-v)} \!+\! i\ve \right) \right] \right\} \! ,
\ee
\ \\[-7mm]
\be \label{Mt2.final}
\wi R_2^{(1)} = \sqrt{z} \left\{ -\fm{1}{4\e} \!+\! \fm{4\pi^2-51}{24} 
+ \e \left[ \fm{6\pi^2-171+80\zeta(3)}{16} 
+ \!\! \int_0^1 \! dx \!\! \int_0^1 \! dy \left( 
 \fm{y-3}{2-2y} {\rm Li}_2(w_1) \!
+\fm{y(3-2y)}{1-y} {\rm Li}_2(w_2)
\right) \right] \right\} \! ,
\ee
\ \\[-6mm]
\be \label{Mt3.final}
\wi{R}_3^{(n)} = \sqrt z \left\{ \begin{array}{ll}
\! \f{n-2}{4 \e} - \! \f{n-2}{2}\ln z - \! \f{n}{8} + \! 1 + \! z
- \! \f{2z+1}{2} \sqrt{1\! -\! 4z} \; f(z) + z (1\! -\! z) f(z)^2, & $n$~{\rm odd}, \\[4mm]
Y^{(n)} + \int_0^1 dx \int_0^1 dy \int_0^1 dv \left\{ \f{1-2x^2y}{x}  
   \ln\left[z - x (1\! -\! x) v y-i \ve\right]-\f{1}{x} \ln z\right\}, & $n$~{\rm even},
\end{array} \right. 
\ee
\ \\[-6mm]
\be \label{Mt4.final}
\wi{R}_4^{(n)} = \sqrt z \left\{ \begin{array}{ll}
\! \f{n-2}{4 \e} + \! \f{3-n}{2} \ln z - \! \f{n+6}{8}
- \! 2 \int_0^1 dx \int_0^1 dy \int_0^1 dv \;
  x y \ln\f{zv+x(1-x)(1-v)(1-v+vy)}{v}, & $n$~{\rm odd}, \\[4mm]
Y^{(n)} + \int_0^1 dx \int_0^1 dy \int_0^1 dv \left\{ \f{1-2x^2y}{x}
\ln\f{zv+x(1-x)(1-v)(1-v+vy)}{v}-\f{1}{x} \ln z \right\}, & $n$~{\rm even},
\end{array} \right. \hspace{-2mm}
\ee
\ \\[-6mm]
\be \label{Mt5.final}
\wi R_5^{(1)} = \f{\sqrt z}{4} \left[ \f{1}{\e} - 2 \ln z - \f{5}{2} + 
\e \left( \f{3}{4} - \f{\pi^2}{6} + 5 \ln z + 2 \ln^2 z \right) \right],
\ee 
\mathindent1cm
where $Y^{(n)} = -\f{1}{2\e^2}+\f{1}{\e} \left(\ln z+\f{1}{4}\right)
-\ln^2 z+\f{\pi^2}{12}-\f{n^2-5n+6}{16}$~ and ~$w_k =
\f{(y-1)z}{x(1-x)y^k}$. The variables $u_k$ and the function $f(z)$
are as in $b(z)$ in eq.~(\ref{bzz}).

We have presented $\wi{R}_3^{(n)}$ and $\wi{R}_4^{(n)}$ for arbitrary
$n$, while ~$n=1$~ was assumed otherwise. In those cases, results for
other ~$n$~ can be found from the following simple relations:
\bea
G_k^{(n+1)} &=& \left\{ \begin{array}{lll} 
\f{-1+\e}{2} \; G_k^{(n)}, & {\rm ~for~} k=1,2, & \hspace{1cm} ({\rm because~~} 
\gamma_{\alpha} \gamma_{\mu} \gamma^{\alpha} = (-2+2\e) \gamma_{\mu}),\\[2mm]
\f{+1+\e}{2} \; G_k^{(n)}, & {\rm ~for~} k=3,4, & 
\hspace{1cm} \mbox{(see eq.~(\ref{gJg}))},
\end{array} \right. \\[2mm]
\wi{G}_k^{(n+1)} &=& ~-\fm{\e}{2} \; \wi{G}_k^{(n)}, {\rm ~~~~~~for~} k=1,2,5,
\hspace{9mm} ({\rm because~~} 
\gamma_{\alpha} \sigma_{\mu\nu} \gamma^{\alpha} = -2\e\, \sigma_{\mu\nu}).
\label{rele}
\eea
The overall factor of $\e$ in the r.h.s. of the latter equation is the
reason why we have calculated the ${\cal O}(\e)$ parts of
$\wi{R}_1^{(1)}$, $\wi{R}_2^{(1)}$ and $\wi{R}_5^{(1)}$. They are
necessary for the ${\cal O}(1)$ matrix elements of $\wi{Q}^{(0)}$,
i.e. they may matter beyond SM.

For the color-octet analogues of $Q^{(n)}$ and $\wi{Q}^{(n)}$, the
expressions for $R_k^{(n)}$ and $\wi{R}_k^{(n)}$ get multiplied by
additional color factors of ~$-\f{1}{6}$~ (for $k \leq 4$) and
~$\f{4}{3}$~ (otherwise). Of course, the values of $R_k^{(n)}$ and
$\wi{R}_k^{(n)}$ remain the same for the mirror copies of $Q^{(n)}$
and $\wi{Q}^{(n)}$, provided $P_R$ is replaced by $P_L$ in
eq.~(\ref{Gk.n}).

Thus, apart from the trivial case of $\wi{G}_6^{(n)}$, the present
section contains complete results for the unrenormalized two-loop
$b\to s \gamma$ diagrams with insertions of four-quark operators, in
the SM and beyond.\footnote{
Up to operators that vanish in 4 spacetime dimensions, i.e. evanescent
operators --- see section~\ref{sec:renor}.}

\newsection{The $\overline{\rm MS}$-renormalized amplitude in the SM}
\label{sec:renor}

In the present section, we shall restrict ourselves to the SM
operators $P_k$ (\ref{ops}) and use
eqs.~(\ref{M1.final})--(\ref{Mt5.final}) to evaluate the
$\overline{\rm MS}$-renormalized on-shell $b\to s\gamma$ amplitude at
the NLO in QCD. From this amplitude one reads out the coefficients
$r_k$, i.e. our main results presented in section~\ref{sec:res}.

The renormalized effective Lagrangian reads\footnote{
Note that the interaction terms in eq.~(\ref{hamil}) have opposite sign.}
\mathindent0cm
\bea
{\cal L}_{\mathrm{eff}} &=& {\cal L}_{\scs \mathrm{QCD} \times \mathrm{QED}}
+ \f{4 G_F}{\sqrt{2}} V^*_{ts} V_{tb} \left\{ C_7(\mu) Z_{77} Z_{\psi} Z_m P_7
+ C_8(\mu) Z_{\psi} Z_m \left( Z_{88} Z_g Z^{\f{1}{2}}_G P_8 + Z_{87} P_7 \right)
\right. \nonumber\\[2mm] && \left. ~~~
+ \sum_{k=1}^6 C_k(\mu) \left[ Z^2_{\psi} \left( \sum_{j=1}^6 Z_{kj} P_j 
+ \sum_{j=3}^4 Z^E_{kj} E^{(1)}_j \right) 
   + Z_{k7} Z_{\psi} Z_m P_7 \right] \right\} + .... \label{lagr}
\eea
\mathindent1cm
The evanescent operators $E^{(1)}_3$ and $E^{(1)}_4$
are as in the appendix of ref.~\cite{Chetyrkin:1998gb}
\be \label{expl.evan.1}
\begin{array}{rl}
E^{(1)}_3 =& (\bar{s}_L \gamma_{\mu_1}
                      \gamma_{\mu_2}
                      \gamma_{\mu_3}
                      \gamma_{\mu_4}
                      \gamma_{\mu_5}     b_L)\sum_q(\bar{q} \gamma^{\mu_1} 
                                                            \gamma^{\mu_2}
                                                            \gamma^{\mu_3}
                                                            \gamma^{\mu_4}
                                                            \gamma^{\mu_5}     q) 
-20 P_5 + 64 P_3, \\[2mm]
E^{(1)}_4 =& (\bar{s}_L \gamma_{\mu_1}
                      \gamma_{\mu_2}
                      \gamma_{\mu_3}
                      \gamma_{\mu_4}
                      \gamma_{\mu_5} T^a b_L)\sum_q(\bar{q} \gamma^{\mu_1} 
                                                            \gamma^{\mu_2}
                                                            \gamma^{\mu_3}
                                                            \gamma^{\mu_4}
                                                            \gamma^{\mu_5} T^a q) 
-20 P_6 + 64 P_4. \end{array}
\ee

The dots in eq.~(\ref{lagr}) stand for other evanescent operators that
do not affect the NLO matrix elements. They are relevant for the NLO
anomalous dimension matrix though.

Below, in the calculation of the $\overline{\rm MS}$-renormalized
amplitude, we shall use the MS-scheme renormalization constants, and
implicitly assume that all the unrenormalized $l$-loop matrix elements
are multiplied by $(4\pi)^{-l\e} e^{l\e\gamma}$. The light ($u, d, s$)
quark masses will be neglected.

In the NLO calculation of $b\to s\gamma$ matrix elements, the
renormalization constants $Z_g$ and $Z_G$ can be set to unity. As far
as $Z_m$ and $Z_{\psi}$ are concerned, we need the one-loop
expressions \linebreak $Z_m = 1-\f{\al}{\pi\e}$~ and 
~$Z_{\psi} = 1-\f{\al}{3\pi\e}$.

The renormalization constants $Z^E_{k3}$ and $Z^E_{k4}$ can be found 
from eqs.~(20)\footnote{
In that equation, the expression ``(counterterms due to
$Z_{\psi}^2$)'' has been missed.}
and (45) of ref.~\cite{Chetyrkin:1998gb}. The only non-vanishing contributions to
those constants at order ${\cal O}(\al)$ are $Z^E_{54} = \f{\al}{4\pi\e}$,
\linebreak $Z^E_{63} = \f{\al}{18\pi\e}$ and $Z^E_{64} = \f{5\al}{48\pi\e}$.

The renormalization constants $Z_{kj}$ can be found from the anomalous
dimension matrix $\hat{\gamma}^{(0)\mathrm{eff}}$ given in eq.~(8) of
ref.~\cite{Chetyrkin:1997vx}\footnote{
Signs of ${\gamma}^{(0)\mathrm{eff}}_{k7}$ and ${\gamma}^{(0)\mathrm{eff}}_{k8}$ are
fixed by our sign convention inside $D_{\mu} \psi = \left(
\partial_{\mu} + i g G_{\mu}^a T^a + i e Q A_{\mu} \right) \psi$.}
\be \label{gamma0}
\hat{\gamma}^{(0)\mathrm{eff}} = \left[
\begin{array}{cccccccc}
\vspace{0.2cm}
-4 & \f{8}{3} &       0     &   -\f{2}{9} &      0    &     0     & -\f{208}{243} &  \f{173}{162} \\ 
\vspace{0.2cm}
12 &     0    &       0     &    \f{4}{3} &      0    &     0     &   \f{416}{81} &    \f{70}{27} \\ 
\vspace{0.2cm}
 0 &     0    &       0     &  -\f{52}{3} &      0    &     2     &  -\f{176}{81} &    \f{14}{27} \\ 
\vspace{0.2cm}
 0 &     0    &  -\f{40}{9} & -\f{100}{9} &  \f{4}{9} &  \f{5}{6} & -\f{152}{243} & -\f{587}{162} \\ 
\vspace{0.2cm}
 0 &     0    &       0     & -\f{256}{3} &      0    &    20     & -\f{6272}{81} &  \f{6596}{27} \\ 
\vspace{0.2cm}
 0 &     0    & -\f{256}{9} &   \f{56}{9} & \f{40}{9} & -\f{2}{3} & \f{4624}{243} &  \f{4772}{81} \\ 
\vspace{0.2cm}
 0 &     0    &       0     &       0     &      0    &     0     &     \f{32}{3} &        0      \\ 
\vspace{0.2cm}
 0 &     0    &       0     &       0     &      0    &     0     &    -\f{32}{9} &     \f{28}{3} \\
\end{array} \right]. \ee
The relation $Z_{kj} = \delta_{kj} + \f{\al}{8\pi\e}
\gamma^{(0)\mathrm{eff}}_{kj}$ holds for all the $Z_{kj}$ except $\left(
Z_{k7}, Z_{k8} \right)_{k=1,...,6}$.  In the latter case, 
\be
Z_{k7} = \f{\al}{16\pi\e} \, \gamma^{(0)}_{k7} \;
       - \; \f{1}{2} \sum_{j=3}^4 Z^E_{kj} \, y^E_j,
\ee
where $\hat{\gamma}$ is related to $\hat{\gamma}^{\mathrm{eff}}$ 
by~\cite{Buras:1994xp}
\be \label{def.geff}
\gamma^{\mathrm{eff}}_{ji} = \left\{ \begin{array}{cc}
\gamma_{j7} +
\sum_{k=1}^6 y_k\gamma_{jk} -y_j\gamma_{77} -z_j\gamma_{87},
& \mbox{when $i=7$ and $j=1,...,6$},\\
\gamma_{j8} +
\sum_{k=1}^6 z_k\gamma_{jk} -z_j\gamma_{88},
& \mbox{when $i=8$ and $j=1,...,6$},\\
\gamma_{ji}, & \mbox{otherwise},
\end{array}
\right.
\ee
with $y_i$ and $z_i$ given in eq.~(\ref{yyzz}).  The numbers
$y_k$ and $y^E_j$ parametrize the unrenormalized on-shell
one-loop $b\to s\gamma$ matrix elements of the 4-quark operators
\bea \label{pk1loop}
\me{P_k}_{\rm 1\;loop} &=& \left(\f{\mu}{m_b}\right)^{2\e} \wi{y}_k 
                                \me{P_7}_{\rm tree} + {\cal O}(\e^2),
\hspace{2cm} (y_k = \lim_{\e\to 0} \wi{y}_k),\\
\me{E^{(1)}_j}_{\rm 1\;loop} &=& \left(\f{\mu}{m_b}\right)^{2\e} \wi{y}^E_j 
                                \me{P_7}_{\rm tree} + {\cal O}(\e^2),
\hspace{2cm} (y^E_j = \lim_{\e\to0} \wi{y}^E_j).\label{ej1loop}
\eea
An easy one-loop calculation gives $\wi{y}_1 = \wi{y}_2 = 0$ and
\be \label{ytk}
\begin{array}{llll}
\hspace{1cm} & \wi{y}_3 = Q_d = -\f{1}{3}, 
              & \hspace{2cm} & \wi{y}_4 = Q_d C_F = -\f{4}{9},\\[2mm] 
& \wi{y}_5 = 4(5-3\e)Q_d, && \wi{y}_6 = 4(5-3\e)Q_d C_F,\\[2mm]
& \wi{y}^E_3 = 16(4-25\e)Q_d, && \wi{y}^E_4 = 16(4-25\e)Q_d C_F.
\end{array}
\ee

The unrenormalized one-loop matrix elements of 
$P_7$ and $P_8$ can be expressed as follows:
\bea
\me{P_7}_{\rm 1\;loop} &=& \f{\al}{4\pi} \; \wi{r}_7 \; \me{P_7}_{\rm tree},\\[2mm]
\me{P_8}_{\rm 1\;loop} &=& 
\left(\f{\mu}{m_b}\right)^{2\e} \left( -Z_{87} + \f{\al}{4\pi} r_8 \right) 
                            \me{P_7}_{\rm tree} + {\cal O}(\al^2,\e).
\eea
No explicit expression for $\wi{r}_7$ is necessary for our purpose. It
is enough to mention that $\wi{r}_7$ is UV-finite (note that 
$Z_{77} Z_{\psi} Z_m = 1 + {\cal O}(\al^2)$), but it contains an IR
divergence. We assume that this IR divergence is regulated with a
small gluon mass.

The quantity $r_8$ reads~\cite{Greub:1996tg}
\be
r_8 = \f{44}{9} - \f{8}{27} \pi^2 + \f{8}{9} i \pi.
\ee
We have verified this result during the calculation of
$\wi{R}_1^{(1)}$ and $\wi{R}_2^{(1)}$.

In the following, we shall need two unrenormalized matrix elements of
evanescent operators
\bea \label{ee4}
\me{P^{c\bar{c}}_4 + \fm{1}{6} P^{c\bar{c}}_3 - \fm{1}{2} Q^{(1)} 
+ \fm{1}{4} Q^{(0)}}_{\rm 2\;loop} &=&
\f{\al}{4\pi} \; e_4 \; \me{P_7}_{\rm tree}+{\cal O}(\e),\\
\me{P^{c\bar{c}}_6 + \fm{1}{6} P^{c\bar{c}}_5 - 8 Q^{(1)} + Q^{(0)}}_{\rm 2\;loop} &=&
\f{\al}{4\pi} \; e_6 \; \me{P_7}_{\rm tree}+{\cal O}(\e), \label{ee6}
\eea 
where $P^{c\bar{c}}_k$ are just the $c\bar{c}$--parts of the
respective $P_k$.  Note that ~$\me{P^{c\bar{c}}_3}_{\rm 2\;loop} =
\me{P^{c\bar{c}}_5}_{\rm 2\;loop} =0$~ due to color factors.  We find
~$e_4 = Q_u-\f{8}{27}Q_d$~ and ~$e_6 = 16Q_u-\f{32}{27}Q_d$.  These
results are independent of $m_c$. Thus, after appropriate replacement
of quark charges, they hold for any flavor circulating in the
loop. Consequently, we can use eqs.~(\ref{ee4}) and (\ref{ee6}) to
express all the two-loop diagrams with Dirac traces in $\me{P_k}_{\rm
2\;loop}$ as linear combinations of the results from
section~\ref{sec:expr}, up to the following additive constants
obtained from $e_4$ and $e_6$:
\mathindent0cm
\be \label{add}
D_4 = u Q_u + d Q_d-\f{8}{27}(u+d)Q_d,
{\rm~~~~~~and~~~~~~}
D_6 = 16(u Q_u + d Q_d)-\f{32}{27}(u+d)Q_d,
\ee
\mathindent1cm
where $u=2$ and $d=3$ are the numbers of active up and down flavors,
respectively.

Such an approach is much more convenient than calculating diagrams
with Dirac traces separately, because $e_4$ and $e_6$ can be (and are)
found using expansion in external momenta and computer algebra, as 
in the so-called matching computations (see e.g. ref.~\cite{expmom}).
More precisely, such an expansion is applied to diagrams with
subtracted subdivergences, because subtracted matrix elements of
evanescent operators must be local (i.e. polynomial in external
momenta). Once they are found, we add the one-loop subdivergences
again, setting the external momenta on shell. Then, the only relevant
subdivergence is proportional to $\me{P_4}_{\rm 1\;loop}$ that we
already know.

Now, we are ready to calculate the unrenormalized two-loop matrix
elements of $P_k$. Let us write them as follows:
\be \label{2.unrenor}
\me{P_k}_{\rm 2\;loop} = \f{\al}{4\pi} \left(\f{\mu}{m_b}\right)^{4\e} 
( x_k + \wi{r}_7 \wi{y}_k ) \me{P_7}_{\rm tree} + {\cal O}(\e),
\ee
where the terms proportional to $\wi{r}_7$ originate from
$\wi{G}_6^{(n)}$ diagrams. Evaluation of $x_k$ amounts to forming
appropriate linear combinations of
eqs.~(\ref{M1.final})--(\ref{Mt5.final}) with $z \to 0$, $z$ or 1, according
to the Dirac and flavor structure of $P_k$. Explicitly
\mathindent0cm
\bea
x_1 &=& -\f{1}{6} \left[ Q_d H_{12}^{(1)}(z) + Q_u H_{34}^{(1)}(z) \right],  \nnb\\[2mm]
x_2 &=& \hspace{9mm} Q_d H_{12}^{(1)}(z) + Q_u H_{34}^{(1)}(z), \nnb\\[2mm]
x_3 &=& Q_d \left[ H_{14}^{(1)}(0)  + H_{14}^{(1)}(1)  
             + \wi{H}_{15}^{(1)} \right],\nnb\\[2mm]
x_4 &=& D_4 + Q_u  \left[ H'_{34}(0) + H'_{34}(z) \right] 
            + Q_d  \left[ 3 H'_{12}(0) + 2 H'_{34}(0) + H'_{12}(z) + H'_{14}(1) 
\right. \nnb\\[2mm] && \left. \hspace{3cm}
-\f{1}{6} \left( H_{14}^{(1)}(0) +  H_{14}^{(1)}(1) + \wi{H}_{14}^{(1)} \right)  
                        + C_F Q_d \wi{H}_{55}^{(1)} \right],\nnb\\[2mm]
x_5 &=& Q_d \left[ H_{14}^{(3)}(0)  + H_{14}^{(3)}(1)  
             + \wi{H}_{15}^{(3)} \right],\nnb\\[2mm]
x_6 &=& D_6 + Q_u  \left[ H''_{34}(0) + H''_{34}(z) \right] 
            + Q_d  \left[ 3 H''_{12}(0) + 2 H''_{34}(0) + H''_{12}(z) + H''_{14}(1) 
\right. \nnb\\[2mm] && \left. \hspace{3cm}
-\f{1}{6} \left( H_{14}^{(3)}(0) +  H_{14}^{(3)}(1) + \wi{H}_{14}^{(3)} \right)  
                        + C_F Q_d \wi{H}_{55}^{(3)} \right],
\eea
\mathindent1cm
where 
\bea
H_{ij}^{(n)}(y) &=& 4^n C_F \sum_{k=i}^j S_k^{(n)}(y), \hspace{28mm} 
\wi{H}_{ij}^{(n)} = 4^n C_F \sum_{k=i}^j \wi{S}_k^{(n)}, \nnb\\[2mm]
H'_{ij}(y) &=& \fm{1}{2} H_{ij}^{(1)}(y) - H_{ij}^{(0)}(y), \hspace{2cm} 
H''_{ij}(y) = 8 H_{ij}^{(1)}(y) - 4 H_{ij}^{(0)}(y), \nnb\\[2mm]
S_i^{(n)}(y) &=& \left\{ \begin{array}{ll}
\left(R_i^{(n)}\right)_{z \to y}, & n=0,1,\\[2mm]
\left[ -R_i^{(3)} + (20-12\e) R_i^{(1)}\right]_{z \to y}, & n=3.
\end{array} \right.
\eea
The relations between $\wi{S}_i^{(n)}$ and
$\left(\wi{R}_i^{(n)}\right)_{z \to 1}$ are the same as between
$S_i^{(n)}(1)$ and $\left(R_i^{(n)}\right)_{z \to 1}$.  The necessity of
introducing $S_i^{(n)}$ and $\wi{S}_i^{(n)}$ follows from the fact
that ordering of the Dirac matrices in $P_5$ and $P_6$ is opposite
to the one in $Q^{(3)}$ and $\wi{Q}^{(3)}$.

After all the above substitutions, one finds 
(up to {\scriptsize ${\cal O}$}$(\e)$)
\mathindent0cm
\be
\begin{array}{rcl}
x_1 &=& \f{46}{243 \e} + \f{833}{729} - \f{1}{3} [a(z)+b(z)] + \f{40}{243} i\pi,\\[2mm]
x_2 &=& -\f{92}{81 \e} - \f{1666}{243} + 2 [a(z)+b(z)] - \f{80}{81} i \pi,\\[2mm]
x_3 &=& \f{248}{81 \e} + \f{2932}{243} - \f{8}{27} \pi^2 + \f{8\pi}{3\sqrt{3}} + 
          \f{32}{9} X_b - a(1) + 2 b(1) + \f{128}{81} i \pi,\\[2mm]
x_4 &=& -\f{46}{243 \e} - \f{1031}{729} + \f{4}{81} \pi^2 - \f{4\pi}{9\sqrt{3}} - 
        \f{16}{27} X_b + \f{1}{6} a(1) + \f{5}{3} b(1) + 2 b(z) - \f{184}{243} i \pi,\\[2mm]
x_5 &=& \f{5552}{81 \e} + \f{42424}{243} - \f{160}{27} \pi^2 + \f{32\pi}{3\sqrt{3}} +
         \f{128}{9} X_b - 16 a(1) + 32 b(1) + \f{2336}{81} i \pi,\\[2mm]
x_6 &=& -\f{4588}{243 \e} - \f{14378}{729} + \f{80}{81} \pi^2 - \f{16\pi}{9 \sqrt{3}} -
\f{64}{27} X_b - \f{10}{3} a(1) + \f{44}{3} b(1) + 12 a(z) + 20 b(z) - \f{3016}{243} i\pi.
\end{array} 
\ee
\mathindent1cm
where the constant $X_b$ as well as the functions $a(z)$ and $b(z)$
have been already defined in section~\ref{sec:res}.

Once we have found the unrenormalized matrix elements of $P_1$, ...,
$P_6$, it is straightforward to calculate the renormalized ones. The
$\overline{\rm MS}$-renormalized amplitude for $b\to s\gamma$ decay 
reads\footnote{
Up to an overall normalization factor of ~$-\f{4 G_F}{\sqrt{2}} V^*_{ts} V_{tb}$.}
\bea 
M &=& \left[ C_7^{(0)}(\mu) + \f{\al}{4\pi} C_7^{(1)}(\mu) \right] 
      \left[ \me{P_7}_{\rm tree}  + \me{P_7}_{\rm 1\;loop} \right] 
+ C_8^{(0)}(\mu) \left[ \me{P_8}_{\rm 1\;loop} + Z_{87} \me{P_7}_{\rm tree} \right] 
\nonumber\\[2mm] && 
+ \sum_{k=1}^6 \left[ C_k^{(0)}(\mu) + \f{\al}{4\pi} C_k^{(1)}(\mu) \right] \left\{ 
\sum_{j=1}^6 \left[ Z_{kj} Z_{\psi} + (1-2\e)(Z_m-1)\delta_{kj} \right] \me{P_j}_{\rm 1\;loop} 
\right. \nonumber\\[2mm] && \hspace{1cm} \left.
+ \me{P_k}_{\rm 2\;loop} + Z_{k7} \me{P_7}_{\rm tree} 
+ \sum_{j=3}^4 Z^E_{kj} \me{E^{(1)}_j}_{\rm 1\;loop} \right\} + {\cal O}(\al^2,\e). 
\label{amplitude}
\eea
When comparing the above equation with eq.~(\ref{lagr}), one should
remember about the identity $Z_{77} Z_{\psi} Z_m = 1 + {\cal
O}(\al^2)$.

The appearance of $Z_{\psi}$ and $Z_m$ in eq.~(\ref{amplitude}) can be
understood without calculating any diagram. It is enough to remember
that insertions of the $Z_{\psi}$-counterterms on internal quark
propagators always cancel with the $Z_{\psi}^{\f{1}{2}}$-counterterms
at the ends of those propagators.  Thus, we are left with a {\em
single} power of $Z_{\psi}$ that corresponds to two external quark
lines. The term proportional to $(Z_m-1)$ is found from
\be
\left[ \me{P_k}_{\rm 1\;loop} \right]_{
m_b(\mu) \to Z_m m_b(\mu)} - \me{P_k}_{\rm 1\;loop},
\ee
where $\me{P_k}_{\rm 1\;loop}$ is given in eq.~(\ref{pk1loop}). 
Remember that $\me{P_7}_{\rm tree}$ depends linearly on $m_b(\mu)$.

When all the matrix elements and renormalization constants are
substituted into eq.~(\ref{amplitude}), one finds that all the
$1/\e$ singularities cancel as they should. Furthermore, when the
effective Wilson coefficients are used, all the logarithms of
$m_b/\mu$ are found to be multiplied by numbers from the 7th column of
$\hat{\gamma}^{(0)\mathrm{eff}}$ (\ref{gamma0}). Thus, the amplitude $M$ takes
the form
\mathindent0cm
\bea
M = \me{P_7}_{\rm tree} \left\{ \left[ 1  + \f{\al}{4\pi} \wi{r}_7 \right] 
\left[ C_7^{(0)\mathrm{eff}} + \f{\al}{4\pi} C_7^{(1)\mathrm{eff}} \right] + \f{\al}{4\pi} 
\sum_{\begin{array}{c}\\[-7mm] {\scs 1\leq k\leq 8}\\[-2mm] {\scs k\neq 7} \end{array}} 
C_k^{(0)\mathrm{eff}} \left[ r_k +\gamma^{(0)\mathrm{eff}}_{k7}\ln\f{m_b}{\mu}\right]\right\}.
\hspace{5mm}
\eea
\mathindent1cm
From the above expression, we have read out the results for $r_k$ that
have been already given in eq.~(\ref{rk}).

\newsection{Conclusions}
\label{sec:concl}

 We have computed the two-loop $b\to s\gamma$ matrix elements of all
 the four quark operators containing no derivatives. This allowed us
 to complete the NLO QCD calculation of $\bar B\to X_s\gamma$ decay by
 including for the first time the two-loop matrix elements of the
 QCD-penguin operators $P_3, ..., P_6$. The values of the
 corresponding parameters $r_k$ that enter the branching ratio are
 collected in table~\ref{tab:rk}. They are also valid in extensions of
 the SM.

 The Wilson coefficients of QCD-penguin operators are small in the
 SM. Consequently, two-loop matrix elements of these operators affect
 the $\bar B\to X_s\gamma$ branching ratio by around $1\%$ only. In
 extensions of the SM, the Wilson coefficients might be larger, which
 would enhance their phenomenological significance. However, it should
 be emphasized that the modification of $C_i$ with $i=3, ..., 6$ would
 not only have a NLO effect (via the two-loop contributions
 evaluated here) but also a LO one (via the effective coefficients
 in eq.~(\ref{ceff})).

 Since in certain extensions of the SM, new operators with different
 Dirac and color structures may contribute, we have performed
 calculations that allow inclusion of such operators if necessary. The
 relevant results can be found in section~\ref{sec:expr}. However, a
 complete NLO analysis in the presence of new operators would require
 extending the 3-loop anomalous dimension computation of
 ref.~\cite{Chetyrkin:1997vx}, which is beyond the scope of the present work.
 
 On the technical side, occurrence of internal b-quark propagators
 required going beyond the techniques developed in
 refs.~\cite{Buras:2001mq,Greub:1996tg} that used expansions in
 $z=m_c^2/m_b^2$. Our exact calculations valid for any $z$ show that
 the expansions performed in refs.~\cite{Buras:2001mq,Greub:1996tg} break down
 around $z \sim 0.5$, and are very accurate up to $z \sim 0.3$,
 i.e. well above the physical value of $z \sim 0.1$.

\newsection{Acknowledgements}

We are grateful to Paolo Gambino for verifying the influence of our
results for $r_k$ on the entries of table~\ref{tab:mag} and on the
final prediction for the branching ratio. J.U. would like to thank
Stefan Fritsch and Michael Spranger for fruitful discussions.  
The work presented here was supported in part by the German
Bundesministerium f\"ur Bildung und Forschung under the contract
05HT1WOA3 and the DFG Project Bu.~706/1-1, and by the Natural Sciences
and Engineering Research Council of Canada.  M.M. was supported by the
Polish Committee for Scientific Research under grant 2~P03B~121~20.

\newpage
\setlength {\baselineskip}{0.2in}
 
\end{document}